\documentclass[runningheads]{llncs}
\usepackage[T1]{fontenc}
\usepackage{graphicx}
\usepackage{hyperref}
\usepackage{color}

\usepackage[inline]{enumitem}
\usepackage{multirow}
\usepackage[normalem]{ulem}
\usepackage{booktabs}
\usepackage{soul,color}
\soulregister\cite7
\soulregister\ref7
\soulregister\pageref7

\useunder{\uline}{\ul}{}
\begin{document}
\title{``To Clean-Code or Not To Clean-Code'' A Survey among Practitioners}
\author{
Kevin Ljung\inst{1}
\and
Javier Gonzalez-Huerta\inst{1}\orcidID{0000-0003-1350-7030}}
\authorrunning{Ljung and Gonzalez-Huerta}
\institute{
Blekinge Institute of Technology, 371 79, Karlskrona, Sweden \\
\email{kevinljung98@gmail.com},
\email{javier.gonzalez.huerta@bth.se}}
\maketitle              % typeset the header of the contribution
\begin{abstract}~\newline

\textbf{Context:} Writing Clean Code understandable by other collaborators has become crucial to enhancing collaboration and productivity. However, very little is known regarding whether developers agree with Clean Code Principles and how they apply them in practice.\\
\textbf{Objectives:} In this work, we investigated how developers perceive Clean Code principles, whether they believe that helps reading, understanding, reusing, and modifying Clean Code, and how they deal with Clean Code in practice.\\
\textbf{Methods:} We conducted a Systematic Literature Review in which we considered 771 research papers to collect Clean Code principles and a survey among 39 practitioners, some of them with more than 20 years of development experience.\\
\textbf{Results:} So far, the results show a shared agreement with Clean Code principles and its potential benefits. They also show that developers tend to write messy code to be refactored later.

\end{abstract}
\section{Introduction}

The development of software systems has turned into a collective endeavour that, in some cases, involves thousands of engineers, distributed globally into hundreds of teams that have to work with code written by others. In this scenario, writing code that is understandable by others becomes crucial. The selection of identifiers, and the length of methods and classes, are, among others, principles that developers should have in mind when writing code.

However, there are no measures to assess code quality universally, and there is a lack of standards for code quality—even the understanding of what code quality is somehow diffuse~\cite{Borstler2017}.

Clean Code \cite{Martin2009} has become one of the most relevant craftsmanship practices for developers worldwide, and several research studies have analyzed its nature and effects. The principles and practices described in the book have been widely embraced as a synonym for code quality by many software developers and software development organizations. However, the evidence of its use in practice, how developers perceive its principles, and how they apply them is scarce in the software engineering literature.

Although there are studies reporting the Clean Code benefits (e.g., \cite{Digkas2020,Koller}), how to support it (e.g.,~\cite{geomar45829}), analyzing challenges and hindrances of its adoption in practice (e.g., \cite{Rachow2018}), or different aspects of refactoring and how it impacts on Clean Code or code quality (e.g.,\cite{Almogahed2018,Almogahed2019,Alomar2021,Ammerlaan2015,Zabardast2020,Arif2020,Jehad2018,Dibble2014,Kim2012,Pantiuchina2020,SaeLim2019,Sharma2015,Vakilian2012}),  it is still unclear how professional developers perceive Clean Code. In this paper, we report a Questionnaire Survey study that explores the practitioners’ perceptions regarding Clean Code. The goal of the survey is to gain an understanding of:
\begin{enumerate*}[label=(\roman*)]
\item the degree of agreement with its principles and practices amongst practitioners; 
\item whether they believe that Clean Code can help them be more efficient and effective while reading, understanding, reusing and maintaining code;
\item and how they deal with Clean Code in their daily work. 
\end{enumerate*}

To gather a more complete list of Clean Code principles and practices, we conducted a Snowballing~\cite{Wohlin2014} Systematic Literature Review using a hybrid method~\cite{Mourao2017}, in which we selected 28 papers in addition to the Clean Code seminal book \cite{Martin2009}.

The remainder of the paper is structured as follows: Section \ref{sec:related} discusses related works in the area. Section~\ref{sec:methodology} describes the details of the Systematic Literature Review and the Questionnaire Survey planning and execution. Section~\ref{sections:results} reports the results of the study. Section~\ref{sec:discussion} discusses the main findings. In Section~\ref{sections:threats} discusses the limitations and threats to the validity. Finally, Section~\ref{sec:conclusions} draws the main conclusions and discusses further works.
\section{Related Work}\label{sec:related}

The Clean Code seminal book somehow refines one of the aspects of Software Craftsmanship, with a deep emphasis on writing high-quality, understandable code, all surrounded by a shared professional culture.

Since then its publication in 2009 there have been several research studies assessing its benefits (e.g., \cite{Digkas2020,Koller}), how to support it (e.g.,~\cite{geomar45829}), analyzing challenges and hindrances to its adoption in practice (e.g., \cite{Rachow2018}), or different aspects of refactoring and how it impacts on Clean Code or code quality (e.g.,\cite{Almogahed2018,Almogahed2019,Alomar2021,Ammerlaan2015,Zabardast2020,Arif2020,Jehad2018,Dibble2014,Kim2012,Pantiuchina2020,SaeLim2019,Sharma2015,Vakilian2012}). 

Several studies also assess what affects code readability, understandability, and maintainability (e.g.,~\cite{Ammerlaan2015,Avidan2017,Dibble2014,Johnson2019,geomar45829,Lee2015,Lerthathairat2011,Sedano2016}) and complexity (e.g.,~\cite{Ajami2019}). B\"orstler et al.~\cite{Borstler2017} also carried out an exploratory study focusing on understanding code quality.

Some other studies, like the ones reported by Stevenson et al.~\cite{Stevenson2018}, or Yamashita and Moonen~\cite{Yamashita2013}, follow a similar methodology, a questionnaire survey study, but with a different focus: code quality aspects or whether developers care about code smells. 

However, we know very little about the practitioners' perceptions of the Clean Code and whether and how they use it in practice~\cite{Rachow2018}. It is still unclear how professional developers perceive Clean Code, whether they agree with its principles and practices, and whether they believe that Clean Code can help them be- ing more efficient and effective while reading, understanding, reusing and maintaining code.

\section{Research Methodology}\label{sec:methodology}

In this paper, we employ two research methods: 
\begin{enumerate*}[label=(\roman*)]
\item a Snoballing Systematic Literature Review (SLR)~\cite{Wohlin2014} with a hybrid search method~\cite{Mourao2017} and 
\item a Questionnaire Survey, following the guidelines by Kitchenham and Pfleeger~\cite{Kitchenham2002}.
\end{enumerate*}

We focus on the following research questions:
\begin{itemize}
\item \textbf{RQ1:} Do developers agree with Clean Code principles?
\item \textbf{RQ2:} Do developers believe that clean code eases the process of reading, understanding, modifying, or reusing code?
\item \textbf{RQ3:} How do developers keep their code ``Clean''?
\end{itemize}

\subsection{Sistematic Literature Review Planning and Execution}

To answer the RQ1, we first wanted to gather a more complete list of clean code principles, going beyond the ones presented by \textit{``Uncle''} Bob Martin in his Clean Code seminal book~\cite{Martin2009}. We conducted the SLR using snowballing~\cite{Wohlin2014}, using a hybrid approach, combining the database search to define the start set with the iterative citations and references analysis (snowballing)~\cite{Mourao2017}. The objective of this SLR is not to describe the \textit{state-of-the-art} regarding the Clean Code, but rather to identify Clean Code principles in addition to the ones presented in the Clean Code book.

We defined the following inclusion criteria\footnote{We applied the abovementioned acceptance criteria to define the start set and during the snowballing iterations.}:
\begin{itemize}
\item Is the paper published in a peer-reviewed English-language journal, conference or workshop proceedings
indexed in Google Scholar?
\item Does the paper include the terms “clean code” or “code quality” in the title, abstract,
or full text?
\item Is the paper published after 2009?
\item Does the paper define principles and practices of clean code or report their usage in
practice?
\end{itemize}
We also excluded papers talking only about static analysis techniques unless there is a strong emphasis on their use in practice.  We opted for excluding papers written before and during 2009 since the Clean Code book was originally published in 2009. The only exception to the criteria above is the Clean Code book, which appears in Table~\ref{table:StagesSLR} as B.

To define the start set (seed), we carried out a database search in March 2021 using Google Scholar with the search string: “clean code” OR “code quality”. The automated search on Google Scholar found 723 papers that were analyzed applying the abovementioned inclusion criteria, which resulted in the inclusion of 9 papers as starting set (designated as S01 to S09 in Table~\ref{table:StagesSLR}).

We then performed four snowballing iterations summarized in Table~\ref{table:StagesSLR} and stopped when we achieved saturation (i.e., we did not find new papers to include), applying the inclusion and exclusion criteria following the process described above, resulting in the inclusion of 18 papers. Each snowballing iteration consisted of backward (i.e., references analysis) and forward snowballing (citations analysis), which improved precision and recall, respectively. In the citation analysis, we found that some papers had hundreds of citations, most of them irrelevant, and therefore we narrowed the scope of the citations inspection to the ones that included ``clean code'' OR ``code quality'', similar to the one used in the start set definition.

\begin{table}[!htpb]
\scriptsize
\begin{tabular}{p{0.15\textwidth}p{0.35\textwidth}p{0.5\textwidth}}
\hline
\textbf{Stage} & \textbf{Citations and References Screened} & \textbf{Papers Included}                     \\\hline
Seed           &                                            & S01\cite{geomar45829}, S02\cite{Rachow2018}, S03\cite{Borstler2017}, S04\cite{Lerthathairat2011}, S05\cite{Digkas2020}, S06\cite{Lerthathairat2011b}, S07\cite{Ammerlaan2015}, S08\cite{Dibble2014}, S09\cite{Lucena2016}  \\\\
Iteration 1    & 23 references and 6 citations              & P1\cite{Stevenson2018}, P2\cite{Steidl2014}, P3\cite{Yamashita2013}, P4\cite{Ajami2019}, P5\cite{Avidan2017}, P6\cite{Lee2015}, P7\cite{Arif2020}, P8\cite{Almogahed2019}, P9\cite{Almogahed2018}, P10\cite{Kim2012}, P11\cite{Jehad2018} \\\\
Iteration 2    & 10 references and 6 citations              & P12\cite{SaeLim2019}, P13\cite{Sharma2015}, P14\cite{Pantiuchina2020}, P15\cite{Vakilian2012}	                       \\\\
Iteration 3    & 0 references and 3 citations               & P16\cite{Alomar2021}, P17\cite{Sedano2016}, P18\cite{Johnson2019}                                \\\\
Iteration 4    & 0 references and 0 citations               &                                              \\\hline
\end{tabular}
\caption{SLR Snowballing Iteration Statistics and Results} \label{table:StagesSLR}
\end{table}

\subsection{Questionnaire Survey Design and Execution}

The Questionaire Survey allowed us to gather developers' opinions regarding Clean Code practices, their benefits, and the way they keep their code ``clean''. The survey was designed following the guidelines by~\cite{Kitchenham2002}. The questionnaire contained a mixture of closed and open questions to understand the participants' views and opinions better. However, for the sake of brevity and clarity, the results presented in this paper focus only on the closed questions.\footnote{The questionnaire is available for download in the companion materials in Zenodo DOI: 10.5281/zenodo.6973656.}

The closed questions in the questionnaire mainly were seven items Likert-scale questions, including a neutral response, which avoids forcing a positive or negative choice, which seems adequate for an exploratory survey. The survey questions were grouped into pages to prevent respondents from being overwhelmed with a long list of questions.

The questionnaire was developed, distributed, and analyzed using \textit{Questback} survey software\footnote{https://www.questback.com}. The questionnaire was distributed by email and using social networks but also spread out to contacts within some companies that redistributed the survey within their respective organizations. Therefore we used non-probabilistic sampling, convenience and snowballing.

Following the guidelines in \cite{Linaker2015} the \textit{Target Audience}, \textit{Unit of Analysis}, \textit{Unit of Analysis}, \textit{Unit of Observation}, and \textit{Search Unit} are software developers with industrial experience, whilst the \textit{Source of Sampling} are the authors' contacts in Swedish and Spanish software industry.

The survey began on April 19th, 2021 and had a programmed end on  May 18th, 2021. The Total Gross Sample was 645 potential invitees. A total of 110 respondents (i.e., $17.05\%$) started the questionnaire (Net Participation). However, only 39 completed the questionnaire ($35.35\%$ of the Net Participation). The completion rate from the Total Gross Sample was $6.05\%$.

We examined the partially completed questionnaires and found out that there was a wide range of cases, but mostly few questions were answered. Therefore we decided to exclude the non-completed questionnaires from the analysis.

For answering RQ1, in which we assess the degree of agreement with the clean code principles, we applied the Wilcoxon test to check if the responses were greater than the neutral value=4 (agreement) to see if the responses were statistically significant.

\section{Results}\label{sections:results}

\subsection{SLR Results: Clean Code Principles}

In Table~\ref{table:PrinciplesSLR} we report the Clean Code principles extracted from the papers included in the SLR. We also list the papers in which principles are mentioned, and whether the papers report evidence of their usage in practice. Most of the principles listed in  Table~\ref{table:PrinciplesSLR} come from the Clean Code book, with one exception: \textit{Minimize Nesting}~\cite{Sedano2016}. The principles listed in  Table~\ref{table:PrinciplesSLR} were the input for the creation of the survey questionnaire questions that aim at answering RQ1.

\begin{table}[!htpb]
\scriptsize
\begin{tabular}{p{0.2\textwidth}p{0.45\textwidth}p{0.35\textwidth}}
\toprule
\textbf{Type}                               & \textbf{Principle}                  & \textbf{Source}\\
\midrule
\multirow{5}{*}{\textbf{General}}           & The Boy Scout Rule                  & B, P02, S05                    \\ \cline{2-3}
                                            & Minimize nesting                             & P17                            \\ \cline{2-3}
                                            & KISS - Keep It Simple, Stupid!               & B, P01, S02                    \\ \cline{2-3}
                                            & OCP - Open Closed Principle                  & B, P01, S06                    \\ \cline{2-3}
                                            & Separate Constructing a System from Its use & B                              \\ \hline
\multirow{6}{*}{\textbf{Naming}}            & Use Meaningful Names                         & B, S01, S02, S04, S06          \\ \cline{2-3}
                                            & Use Intention-Revealing Names                & B                              \\ \cline{2-3}
                                            & Pronounceable Names                          & B                              \\ \cline{2-3}
                                            & Searchable Names                             & B                              \\ \cline{2-3}
                                            & Avoid Disinformation                         & B                              \\ \cline{2-3}
                                            & Avoid Mental Mapping                         & B                              \\ \hline
\multirow{8}{*}{\textbf{\shortstack[l]{Function \\and \\Method}}}       & Do One Thing     &B, P02,S06                      \\ \cline{2-3}
                                            & Command Query Separation                     &B                               \\ \cline{2-3}
                                            & Extract Try-Catch Block                      &B                               \\ \cline{2-3}
                                            & Have No Side Effects                         &B, S06                          \\ \cline{2-3}
                                            & DRY - Don't Repeat Yourself                  &B, S02, S07, P09, P11, P14, P17 \\ \cline{2-3}
                                            & Function Arguments                           &B, S06                          \\ \cline{2-3}
                                            & Structured Programming                       &B                               \\ \cline{2-3}
                                            & Methods/Functions should be small            &B,S06, P01$^*$                  \\ 
\hline
\multirow{6}{*}{\textbf{Comments}}          & Amplification                                &B                               \\ \cline{2-3}
                                            & Clarification                                &B                               \\ \cline{2-3}
                                            & Explain Yourself in Code                     &B                               \\ \cline{2-3}
                                            & Explanation of Intent                        &B                               \\ \cline{2-3}
                                            & TODO Comments                                &B                               \\ \cline{2-3}
                                            & Warning of Consequences                      &B                               \\ \hline
\multirow{5}{*}{\textbf{Formatting}}        & Team Coding Standards                        &B, S03, S08, S09, P17           \\ \cline{2-3}
                                            & Horizontal Formatting - Indentation          &B                               \\ \cline{2-3}
                                            & Dependent Functions                          &B                               \\ \cline{2-3}
                                            & Vertical Distance and Ordering               &B                               \\ \cline{2-3}
                                            & Organizing for Change                        &B                               \\ \hline
\multirow{2}{*}{\textbf{\shortstack[l]{Object and \\Data Structures}}} & Data/Object Anti-Symmetry&B                        \\ \cline{2-3}
                                            & Law of Demeter                               &B                               \\ \hline
\multirow{4}{*}{\textbf{Error Handling}}    & Prefer Exceptions to Returning Error Codes   &B                               \\ \cline{2-3}
                                            & Don't Pass Null                              &B                               \\ \cline{2-3}
                                            & Don't Return Null                            &B                               \\ \cline{2-3}
                                            & Write Your Try-Catch Statement First         &B                               \\ \hline
\multirow{3}{*}{\textbf{Unit Tests}}        & Keeping Tests Clean                          &B                               \\ \cline{2-3}
                                            & One Assert per Test                          &B                               \\ \cline{2-3}
                                            & Single Concept per Test                      &B                               \\ \hline  
\multirow{8}{*}{\textbf{Class}}             & Class Organization                           &B                               \\ \cline{2-3}
                                            & High Cohesion                                &B, S02, S03, S04, S05, S06, S08, P01$^*$, P8, P9, P11, P12, P14, P16 \\ \cline{2-3}
                                            & Low Coupling                                 &B, S02, S03, S04, S05, S06, S08, P1$^*$, P8, P9, P11, P14, P16 \\ \cline{2-3}
                                            & Encapsulation - Separation of Concerns       &B, P11, P14                     \\ \cline{2-3}
                                            & Isolating from Change                        &B                               \\ \cline{2-3}
                                            & SRP - Single Responsibility Principle        &B, S02, S06, S07, S08, P01$^*$, P3, P16\\ \cline{2-3}
                                            & Minimal Classes and Methods                  &B                               \\ \cline{2-3}
                                            & One Level of Abstraction per Function        &B, P4$^*$, P14, P16             \\ \cline{2-3}
                                            & Classes should be small                      &B, S02, S06, P01$^*$            \\
\bottomrule
\multicolumn{3}{l}{$^*$ Reports evidence of the use of the principle in practice.}\\
\end{tabular}
\caption{Clean Code Principles extracted from the SLR, including sources where the principle is mentioned, and wether there is evidence of its usage in practice} \label{table:PrinciplesSLR}
\end{table}

\subsection{Survey Results}\label{sections:results_survey}

\subsubsection{Demographics}
\begin{figure}[!ht]
\includegraphics[width=\textwidth]{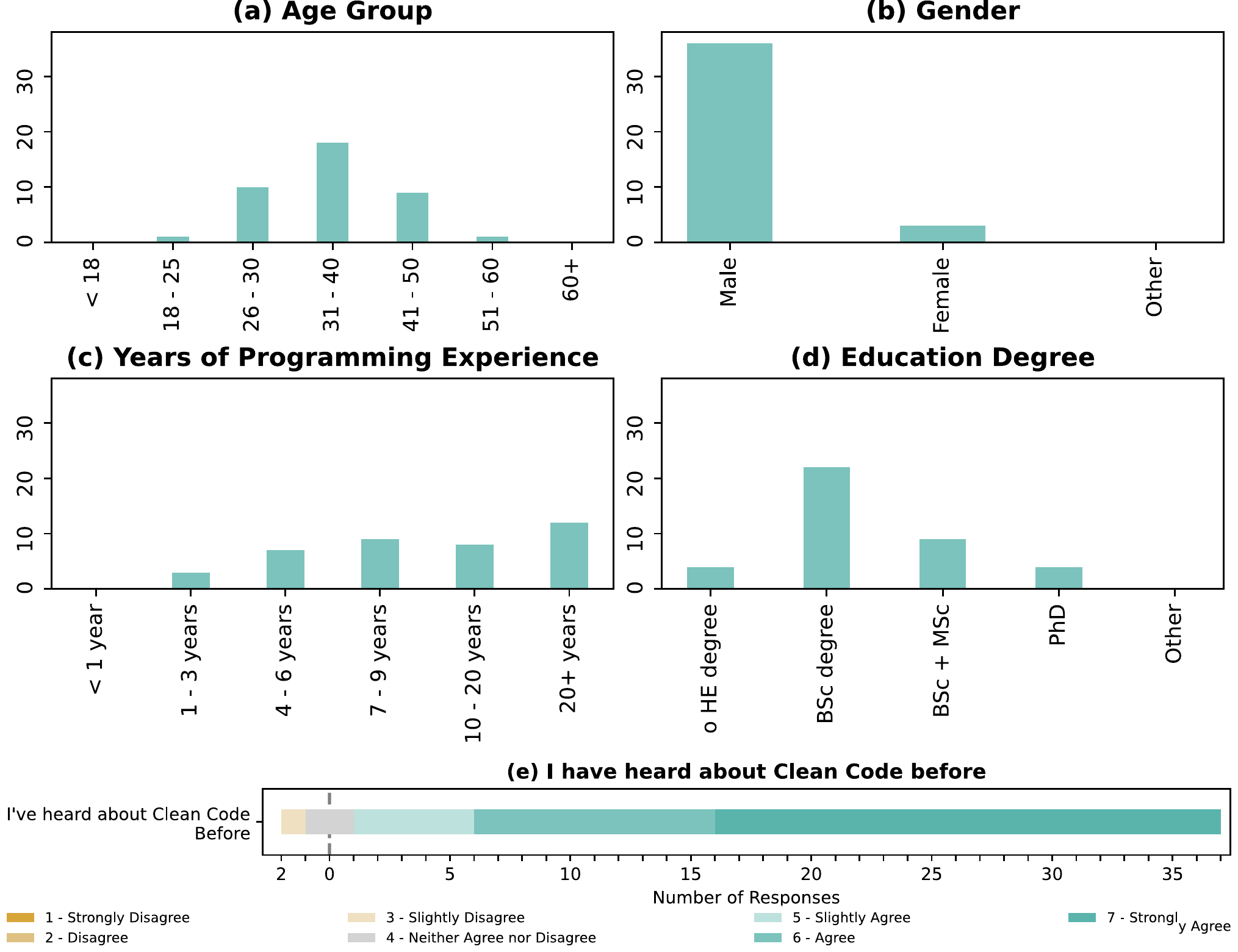}
\caption{Participants' Demographics Including Age Group, Gender, Years of Programming Experience, Education, and Familiarity with Clean Code.} \label{fig:demographics}
\end{figure}

As shown in \ref{fig:demographics}, from the 39 participants that have completed the survey, the majority are in the age group 31-40 (Figure~\ref{fig:demographics}.(a)), 36 are male 3 are female (Figure~\ref{fig:demographics}.(b)), they have more than 20 years of experience (Figure~\ref{fig:demographics}.(c)), and most of them have a BSc degree, and even some have a PhD (Figure~\ref{fig:demographics}.(d)). In addition to that, as shown in Figure~\ref{fig:demographics}.(e), the vast majority of the respondents are familiar with the Clean Code concept.

Based on these demographics, although the number of participants is not very big, we believe the participants constitute a relevant group of respondents for addressing the research questions.

\subsubsection{RQ1: Do developers agree with Clean Code principles?}
\begin{figure}[!ht]
\includegraphics[width=\textwidth]{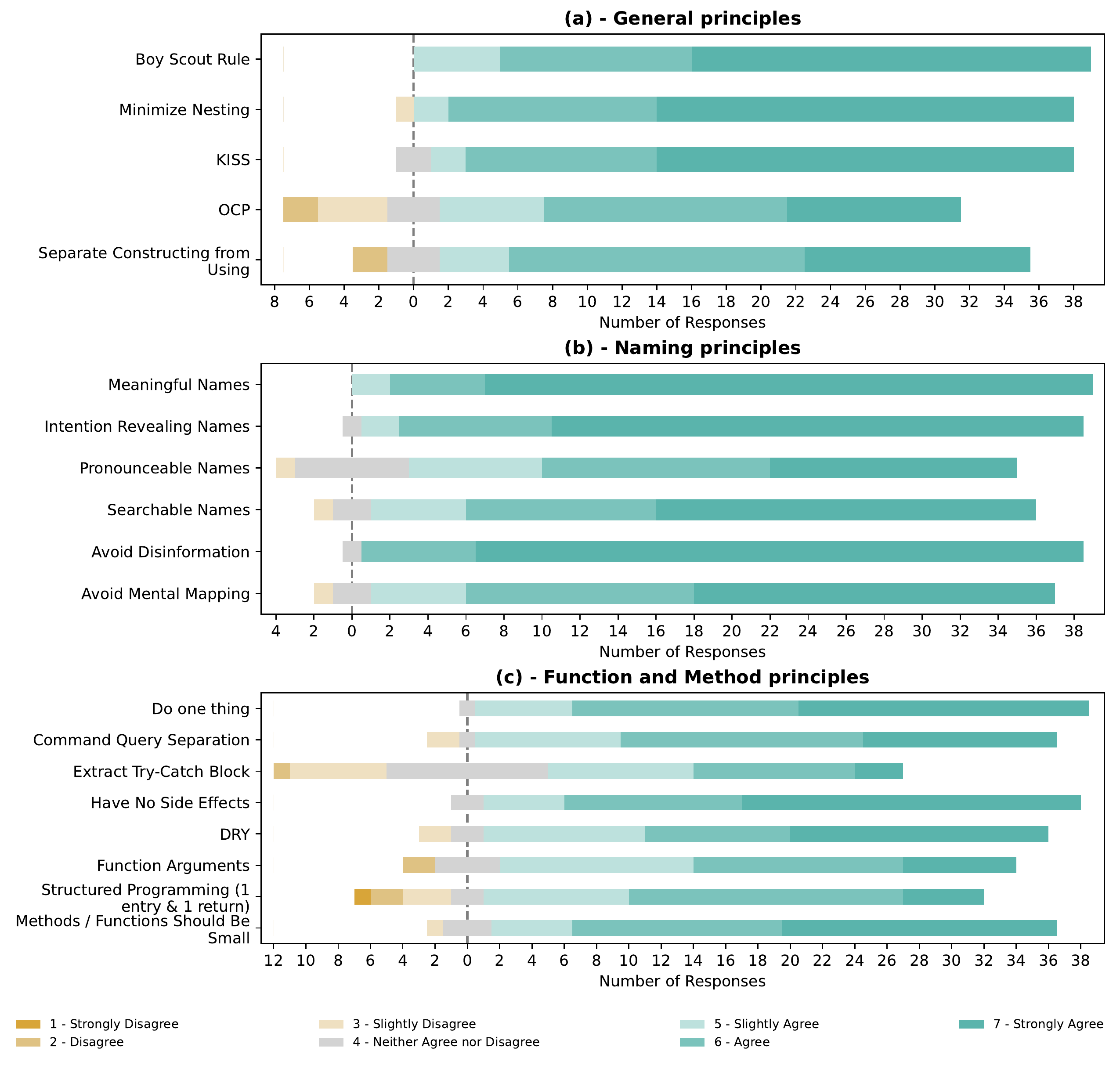}
\caption{Developers' degree of agreement with the General, Naming, and Function, and Method Principles.} \label{fig:principles1}
\end{figure}

Figure~\ref{fig:principles1} shows the developers' degree of agreement with the \textit{General}, \textit{Naming}, and \textit{Function an Method} Principles. The majority of the participants tend to agree with the principles listed, being OCP - Open Closed Principle~\cite{Martin2009} and Extract Try-Catch block~\cite{Martin2009} the most controversial in these groups. The Wilcoxon signed-rank test results ($p-value<0.05$) show that the answers were significantly greater than the neutral value  (i.e., the answers were greater than the neutral value equals 4) for all the principles listed in Figure~\ref{fig:principles1}. These results confirm that the participants agree with these Clean Code principles\footnote{The complete results of the Wilcoxon signed-rank test are available in the companion materials in Zenodo DOI: 10.5281/zenodo.6973656}.

\begin{figure}[!ht]
\includegraphics[width=\textwidth]{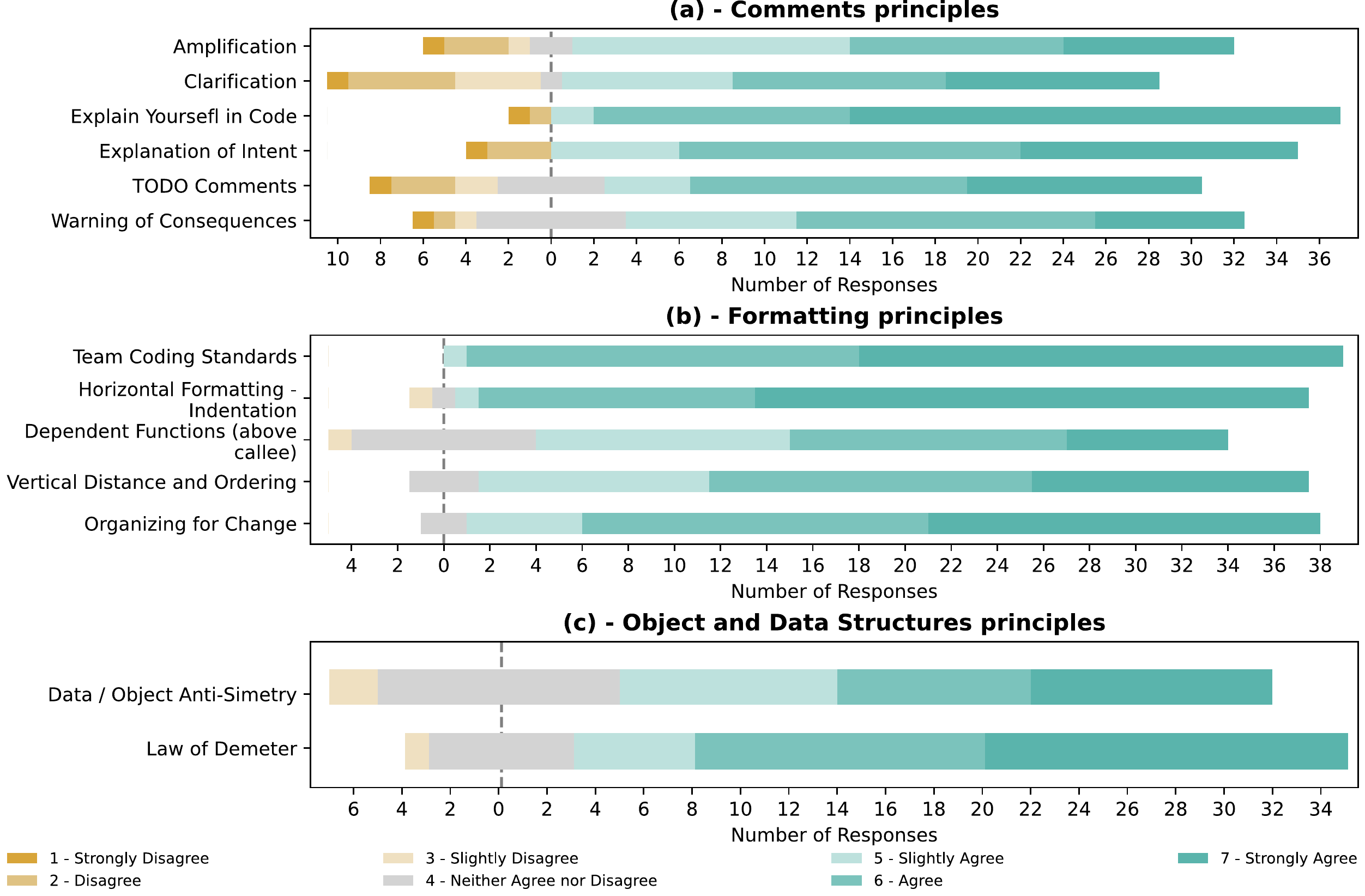}
\caption{Developers' degree of agreement with the Comments, Formatting and object and Data Structures Principles.} \label{fig:principles2}
\end{figure}

Similarly, Figure~\ref{fig:principles2} shows the degree of agreement with the \textit{Comments}, \textit{Formatting} and \textit{Object and Data Structures} Clean Code principles. Although there are some disagreements with the Comments principles. Again, the Wilcoxon signed-rank test results are statistically significant ($p-value<0.05$), confirming that the participants tend to agree with these principles.

\begin{figure}
\includegraphics[width=\textwidth]{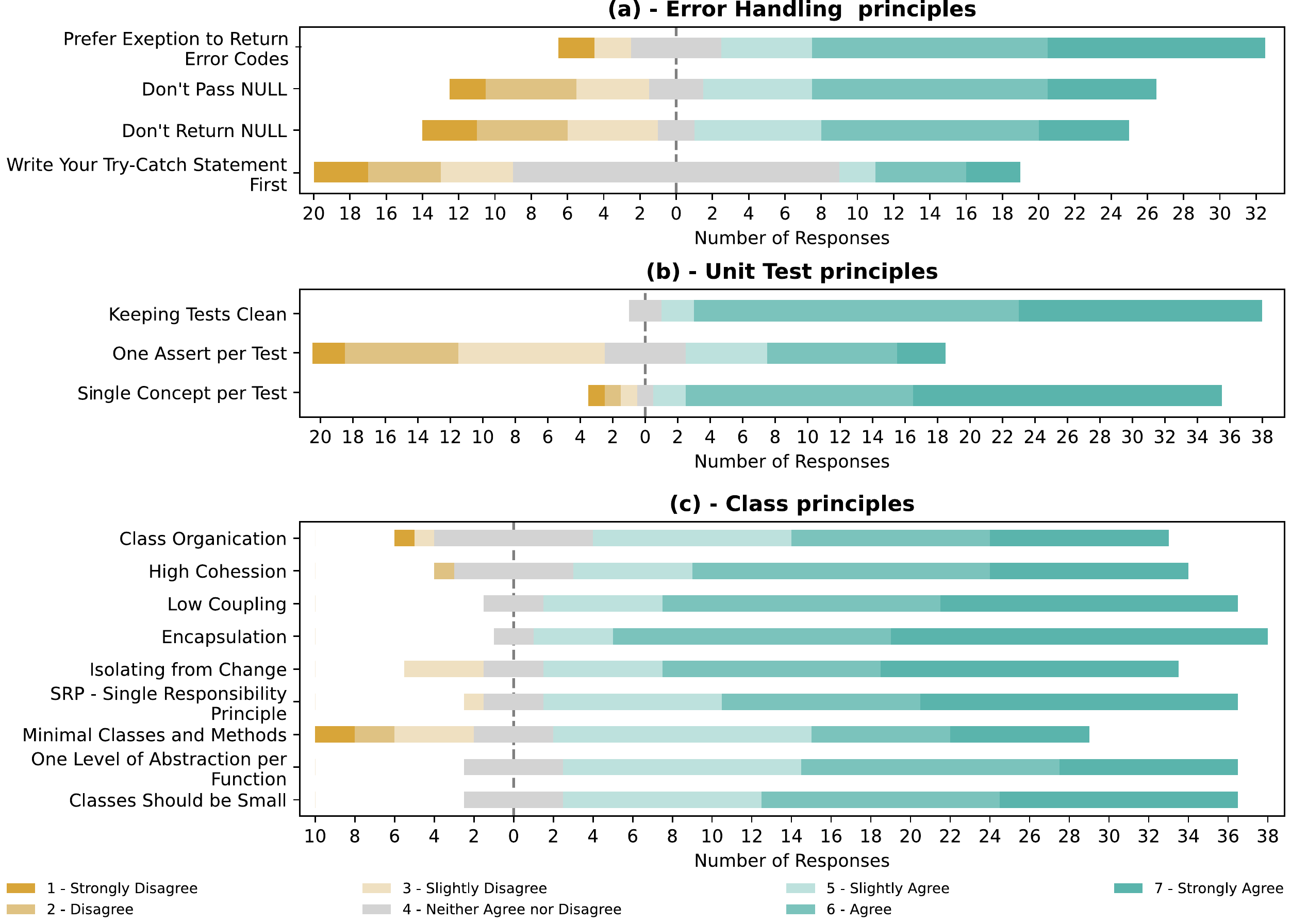}
\caption{Developers' degree of agreement with the Error Handling, Unit Test, and the Class Clean Code principles} \label{fig:principles3}
\end{figure}

Finally,Figure~\ref{fig:principles3} shows the degree of agreement with the Error Handling, Unit Test, and the Class Clean Code principles. In this case, all the principles were statistically significant except two: \textit{Write Your Try-Catch First} ($p-value=0.47$) and \textit{One Assert Per Test} ($p-value=0.41$). Therefore we can conclude that developers agree with the majority of the Clean-Code principles except 2 (\textit{Write Your Try-Catch First} and  \textit{"One Assert Per Test")}.

\subsubsection{RQ2: Do developers believe that clean code eases the process of reading, understanding, modifying, or reusing code?}
\begin{figure}
\includegraphics[width=\textwidth]{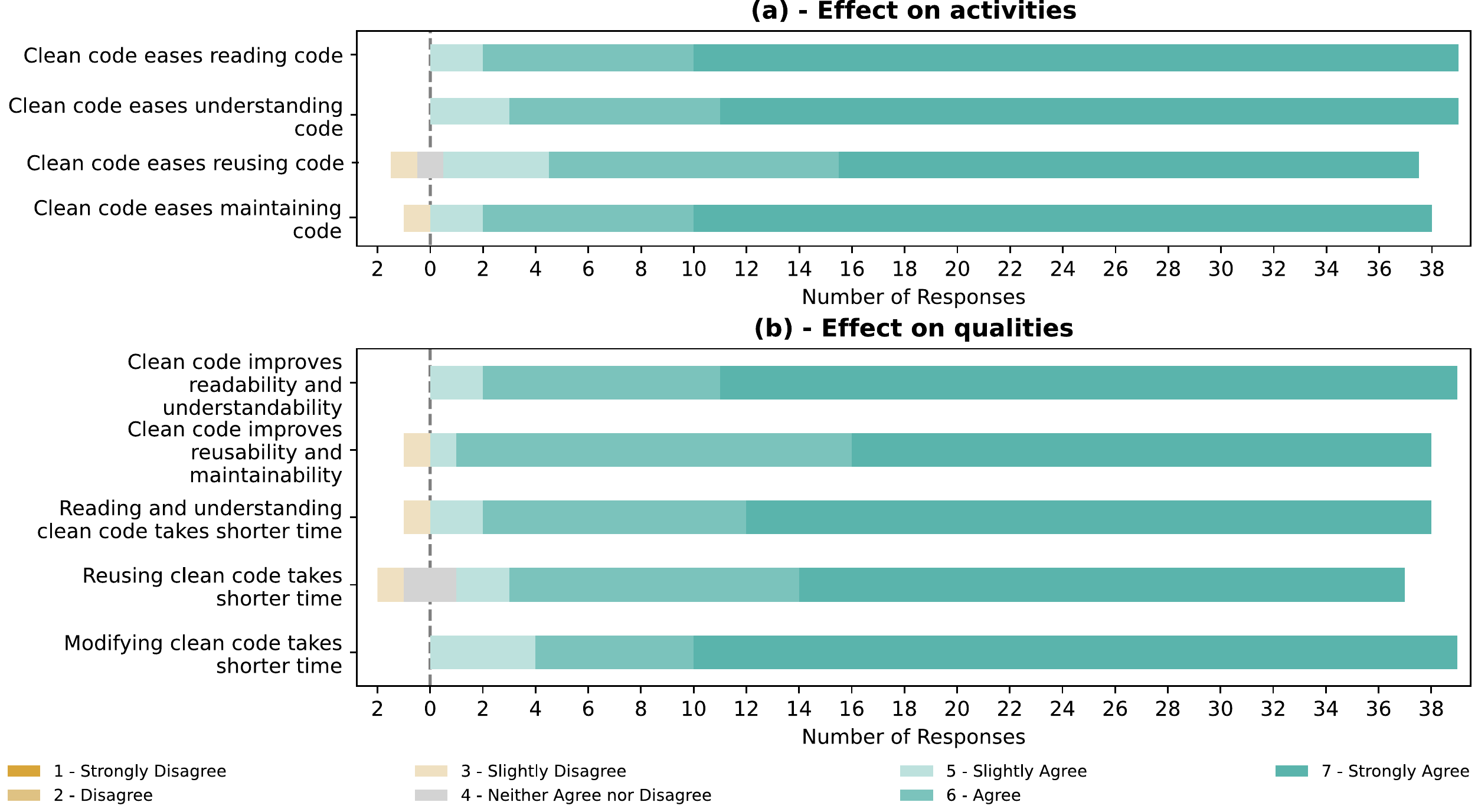}
\caption{Developers' perceptions regarding whether Clean Code eases the tasks of reading, understanding, reusing, and maintaining the code, its impact on readability, understandability, and maintainability, and whether it is faster to interact with Clean Code.} \label{fig_qualities}
\end{figure}

As shown in Figure~\ref{fig_qualities}, participants agree that Clean Code eases the different code-related developer activities, i.e., reading, understanding, reusing and maintaining the code. The participants also believe that Clean Code improves the quality attributes of the code, i.e., understandability, reusability, and maintainability (see Figure~\ref{fig_qualities}.(b)). They also agree that reading, reusing, and modifying Clean Code takes a shorter time than working with ``messy'' code. The Wilcoxon signed-rank test results ($p-value<0.05$) show that the answers were significantly greater than the neutral value for all these questions.

\subsubsection{RQ3: How do developers tackle the process of having Clean Code?}

\begin{figure}

\includegraphics[width=\textwidth]{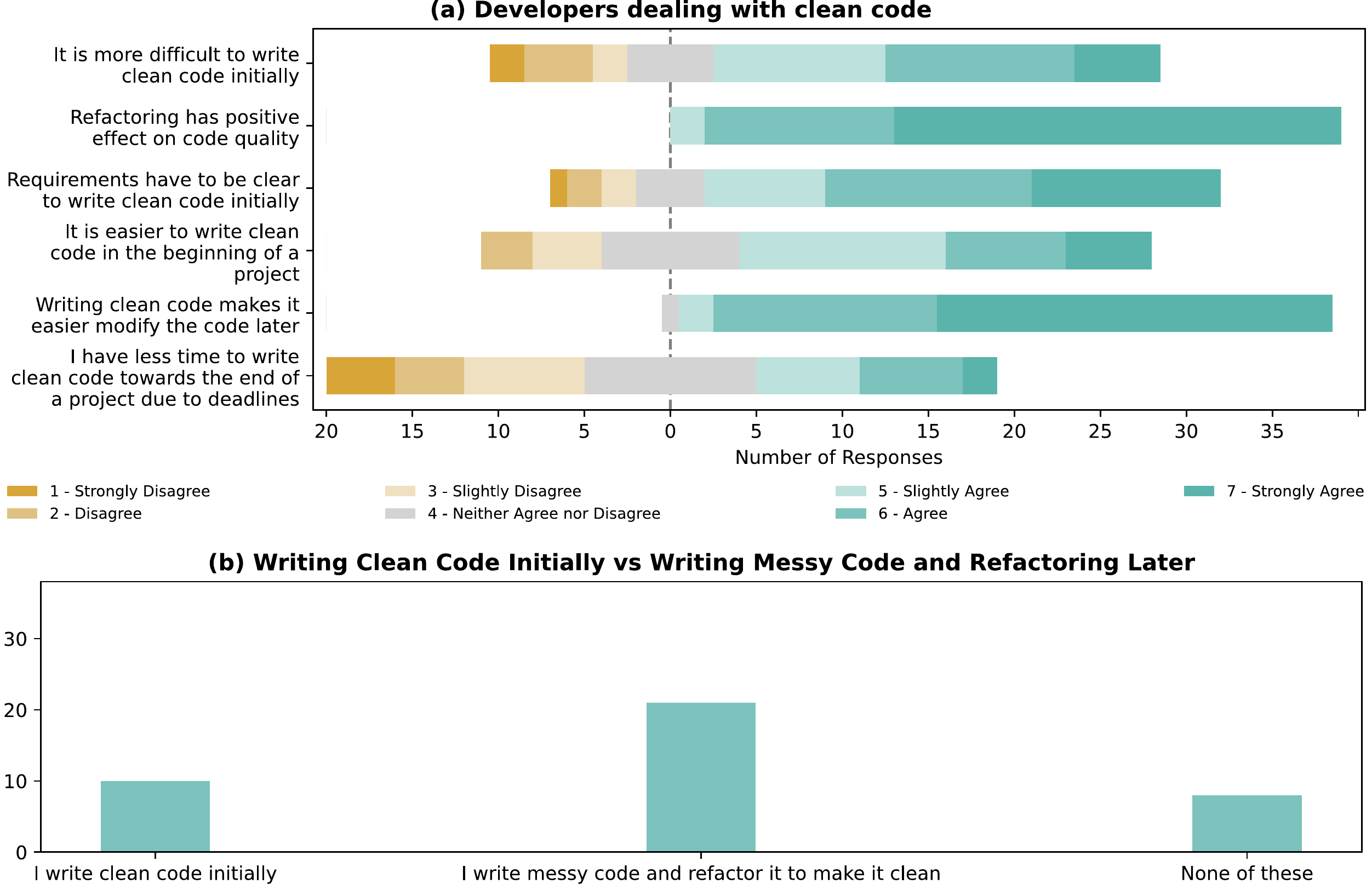}
\caption{Participant's perceptions regarding how they deal with Clean Code.} \label{fig_how_to1}
\end{figure}

Figure~\ref{fig_how_to1} shows the participants' responses to the questions regarding how they deal with Clean Code:
\begin{itemize}
\item whether respondents find it more difficult to write clean code initially as compared to writing ``messy'' code to be later refactored.
\item their perceptions about the impact that refactoring has on code quality.
\item whether they believe that requirements need to be clear to be able to write Clean Code initially.
\item whether they believe that it is easier to write clean code at the beginning (i.e., in early phases) of a project.
\item whether writing clean code makes it easier to modify the code later.
\item whether they perceive they have less time to write clean code towards the end of a project due to deadlines (i.e., time pressure).
\end{itemize}

The participants generally agree with all those statements, but the one states that deadlines prevent them from writing code. Indeed the Wilcoxon signed-rank test results ($p-value<0.05$) show that the answers were significantly greater than the neutral value for all these questions but the last one. In this case, we also ran the test with the symmetric hypothesis (to check whether the disagreement was statistically significant), since the visual inspection of Figure~\ref{fig_how_to1} seems to have more disagreeing answers. However, the results were not statistically significant ($p-value>0.05$).

\section{Discussion}\label{sec:discussion}

\subsubsection{RQ1: Do developers agree with Clean Code principles?}

As reported in Section~\ref{sections:results_survey}, the participants agree with most of the Clean Code principles; there are only two in which the statistical test did not find any agreement: \textit{Write Your Try-Catch First} and \textit{One Assert Per Test}. \textit{Write Your Try-Catch First}  might be controversial since it probably dictates too much about how to do TDD (or one can argue that it forces you even to do it). TDD is not just tests-first or test-last but making sure developers take fine-grained develop-test steps and shorten the feedback loops~\cite{Fucci2015}. One might prefer the exception to be thrown until they have finished writing and testing the actual code. Regarding \textit{One Assert Per Test}, it can also be argued that it can be a source of test clones, and developers might prefer to have several asserts in one test. The \textit{Comments} principles probably generate reluctance since the principles advocate for not commenting on the code but instead writing Clean and `clear'', self-explanatory code. In the open questions, some of the participants refer to the use of comments to clarify what the code does, which contradicts the Clean Code principles. However, most participants tend to agree with the \textit{Comments} principles.

\subsubsection{RQ2: Do developers believe that clean code eases the process of reading, understanding, modifying, or reusing code?}

As shown in Figure~\ref{fig_qualities}, developers believe that Clean Code eases the different code-related developer activities, i.e., reading, understanding, reusing and maintaining the code. The participants also agree with the fact that Clean Code improves the quality attributes in the code (i.e., understandability, reusability, and maintainability). Although there are results that report improvements in maintainability (e.g., \cite{Arif2020}), there are also other studies that report that the impact on understandability is not that obvious. For example, Ammerlan et al.,\cite{Ammerlaan2015} report that understandability can sometimes be hindered when we refactor to clean our code.

We also asked our participants in the questionnaire survey if they believe that it takes a shorter time to read, understand, modify, or reuse clean code compared to unclean code. The respondents strongly agreed with this. Only some developers disagreed that clean code would take a shorter time to read and understand than unclean code. Arif and Rana~\cite{Arif2020} reported that if developers remove code smells in advance and make the code clean, it will take 7\% less effort to add new features to the code than with unclean code. Other research studies (e.g.,~\cite{Koller}) suggest that Clean Code has an impact on the time required to change current functionality, although it does not seem to have an impact on the time used to implement new significant functionality or solve bugs, or to solve small coding tasks. Therefore more research seems to be required to analyze these phenomena in industrial settings.

\subsubsection{RQ3: How do developers tackle the process of having Clean Code?}
As shown in Figure~\ref{fig_how_to1}, participants first acknowledge that they find it more difficult to write Clean Code and that they tend to write ``messy'' code that they refactor later. All participants agree with the fact that refactoring has a positive effect on code quality and, therefore, helps keeping the code clean. Refactoring is probably the most popular technique to keep the code clean and repay Technical Debt~\cite{Digkas2020}. However, it can also negatively affect code quality and introduce Technical Debt Items~\cite{Zabardast2020}. In the open questions, in the question \textit{``What are/would be the challenges with refactoring unclean code to become clean code?''}, some participants mentioned \textit{``Breaking the Functionality''}. However, having enough test coverage might solve that issue since tests are the ``safe net'' when refactoring~\cite{Fowler2018}.

In addition, we also found that developers agreed that the requirements must be specified clearly to write clean code. Only a few developers disagreed with the previous statement. Therefore, the findings from both the literature and the results seem to be aligned. Lucena and Tizzei~\cite{Lucena2016} also mention that integrating all of the previously mentioned methodologies can help the development team write the requirements more precisely. It also showed that the development had a more sustainable velocity and could deliver a more valuable project to the customer when applying these practices.

There was a mixture of opinions regarding whether participants feel they have less time to write clean code towards the end of the project due to deadlines (or when they approach deadlines). These results somehow contrast previous works (e.g., \cite{Rachow2018}), in which time pressure was one of the main reasons for developers not writing clean code.
\section{Threats to Validity}\label{sections:threats}
In this section, we discuss threats to construct, internal, external validity, and reliability.

\noindent \textbf{Construct Validity} concerns mapping the constructs (research questions) to the questions in the survey questionnaire. To illustrate this mapping, we have reported the results and discussion per research question. However, it is still possible that some of the questions do not connect to the construct. 
%One possible solution can be to do a Cronbach's alpha test to check the reliability of the questionnaire. However, in some cases, especially regarding the questions associated with RQ1, they do not necessarily need to be connected. RQ1 is the most exploratory part, and we are trying to establish whether some of the principles are (according to the respondents) connected to Clean Code. 

\noindent \textbf{Internal Validity} is mainly affected by the fact that we derive the findings and conclusions from several questions by `merging' or `abstracting out' the conclusions. This derivation of the results might open the door to researcher bias since we might cherry-pick and give more value to some questions. We have tried to minimize this threat by describing and analyzing the questions separately and also using the Wilcoxon signed-rank test to test the significance of each variable individually. The selection of participants in this study is another threat to the internal validity. The survey was shared through social networks (LinkedIn, Twitter) and the authors' industrial contacts. This publication strategy translates into a limited selection procedure. In addition to that, more than half of the participants did not finish the questionnaire (we discarded their responses), which also represents a self-selection protocol, and might also bias the results. Respondents with more negative views toward Clean Code or less mainstream opinions might have left the questionnaire unfinished. Finally, there might be a tendency to respond with best practices instead of reporting bad behaviours. Once distributed, the researchers had very little control over the questionnaire, which had only a closing deadline, which somehow removed some validity threats. 

\noindent \textbf{Internal Validity} is concerned with the potential generalizability of the results. The number of respondents compared to the potential target population (the software developers working in the industry around the world) is minimal. However, it is in line with similar studies in the area. The question is whether the sample is representative of the \textit{Target Audience}~\cite{Kitchenham2002}. It seems that the respondents' experience is higher as compared to other studies (e.g., \cite{Stevenson2018}). 

\noindent \textbf{reliability} to enhance reliability, we have published the answers to multiple choice questions and the results of the statistical test as companion materials in Zenodo. We keep the answers to the open questions anonymous since, in some cases, those can help identify the respondents' affiliations, and that would break the anonymity clauses on the pre-questionnaire consent form.
\section{Conclusions and Further Work}\label{sec:conclusions}

This paper reports a Questionnaire Survey with $39$ participants that explores how professional developers perceive Clean Code. 
To collect Clean Code principles beyond the seminal book, we conducted a Snowballing Systematic Literature Review using a hybrid search strategy. We inspected $771$ research papers: $723$ papers to define the start set and $48$ papers in the four snowballing iterations. The  Systematic Literature Review resulted in including in the questionnaire one Clean Code principle not listed in the seminal book (i.e., \textit{Minimize Nesting}).

The survey results indicate that developers tend to agree with most of the Clean Code principles, except for two, namely \textit{Write Your Try-Catch First} and \textit{One Assert Per Test}. The results also indicate that developers believe that Clean Code eases reading, understanding, reusing and maintaining the code. They also believe that clean code improves readability, understandability, reusability and maintainability, and that it shortens the time required to read, understand, reuse, and modify the code (in this case, there exists empirical evidence that confirms and disproof these results). The results also indicate that developers tend to write messy code that they refactor later and that refactoring positively affects code quality. They find it more challenging to write clean code initially, that at the beginning of a project seems to be easier to write clean code. There is no consensus on whether time pressure impacts the time they can devote to writing Clean Code. 

These results are the first step toward understanding what principles developers are more prone to adopt and how they try to clean their code. Our results can also help us to understand the support they need and the potential risks they might incur when refactoring. However, these results are only valid in the context of the study, and it is, at this point, difficult to establish generalizations. Therefore there is still a need to conduct similar studies, or replications of this survey in other contexts, even among contributors to Open-Source systems, to strengthen the evidence.

\subsubsection{Acknowledgements} This research was supported by the KKS foundation through the SHADE KKS H\"{o}g project (ref: 20170176) and through the KKS SERT Research Profile project (ref. 2018010) Blekinge Institute of Technology.

 \bibliographystyle{splncs04}
 \bibliography{bibliography}

\end{document}